# Time fractional Schrödinger equation

Mark Naber [a]

Department of Mathematics

Monroe County Community College

Monroe, Michigan, 48161-9746

The Schrödinger equation is considered with the first order time derivative changed to a Caputo fractional derivative, the time fractional Schrödinger equation. The resulting Hamiltonian is found to be non-Hermitian and non-local in time. The resulting wave functions are thus not invariant under time reversal. The time fractional Schrödinger equation is solved for a free particle and for a potential well. Probability and the resulting energy levels are found to increase over time to a limiting value depending on the order of the time derivative. New identities for the Mittag-Leffler function are also found and presented in an appendix.

[a] Electronic mail: mnaber@monroeccc.edu

## I. INTRODUCTION

A Gaussian distribution for random walk problems, in the continuum limit, can be used to generate the ordinary diffusion equation (ignoring boundary conditions and sources)[1]

$$\frac{\partial}{\partial t} U = c \frac{\partial^2}{\partial x^2} U. \tag{1}$$

U represents the concentration of the diffusing material. c is the diffusion coefficient which is positive and whose magnitude helps determine the speed at which the diffusion takes place, and, $t$ and $x$ are the temporal and spatial coordinates. The diffusion coefficient may depend on the coordinates and/or the concentration. When non-Gaussian distributions are used fractional diffusion equations are produced. Diffusion that is generated by non-Gaussian distributions is typically referred to as anomalous diffusion. There is a growing body of examples of phenomena that lie outside of what is predicted by the ordinary diffusion equation (see for example Refs. 2 - 7).

Three types of fractional diffusion equations can be produced when considering non-Gaussian distributions. The first is the space fractional diffusion equation

$$\frac{\partial}{\partial t} U = c \frac{\partial^\beta}{\partial x^\beta} U. \tag{2}$$

Here $0 < \beta \leq 2$ (in this paper Caputo fractional derivatives will be used, the reader is directed to the first appendix for definitions and notation conventions). In this case, the diffusion is still Markovian but only exhibits Brownian motion for $\beta = 2$.

The second type is the time fractional diffusion equation

$$\frac{\partial^\alpha}{\partial t^\alpha} U = c \frac{\partial^2}{\partial x^2} U. \tag{3}$$

In this case the diffusion is non-Markovian and can further be divided into sub- $(0 < \alpha < 1)$ and super- $(1 < \alpha < 2)$ diffusive behavior. Additionally, the mixed case can be considered with both space and time fractional derivatives

$$\frac{\partial^\alpha}{\partial t^\alpha} U = c \frac{\partial^\beta}{\partial x^\beta} U. \tag{4}$$

The Schrödinger equation has the mathematical appearance of a diffusion equation and can be derived by considering probability distributions. Feynman and Hibbs used a Gaussian probability distribution in the space of all possible paths, for a quantum mechanical particle, to derive the Schrödinger equation [8]. Therefore, it is reasonable to consider the different types of Schrödinger equations that are obtainable for non-Gaussian distributions. In an earlier sequence of papers (see Refs. 9 - 11) Laskin constructed space fractional quantum mechanics. This was done using Feynman's path integral approach, the difference being the use of Levy distributions instead of Gaussian distributions for the set of possible paths. The Schrödinger equation that was obtained had space fractional derivatives. This is the same result that one obtains when studying diffusion processes based on Levy distributions instead of Gaussian distributions. Similarly, a time fractional Schrödinger equation would be obtained if one considered non-Markovian evolution. Laskin was able to show that the fractional Hamiltonian was Hermitian and that parity is conserved. Energy levels were also computed for the hydrogen atom and the harmonic oscillator.

In this paper, properties of the time fractional Schrödinger equation are examined. The time fractional Schrödinger equation will be constructed by rewriting the Schrödinger equation so that all derivative operators appear as dimensionless objects, the time derivative is then fractionalized, and the imaginary unit is raised to the order of the fractional time derivative. This last step is important because it ensures the same physical character of the time fractional Schrödinger equation no matter what the order.

For the time fractional Schrödinger equation there are two cases: one for the order of the time derivative being between zero and one and another case for the order being between one and two. For the later case the resulting equation has the same draw backs as the Klein-Gordon equation, in that the initial value of the first derivative and the wave function itself must be specified.

In the following section the time fractional Schrödinger equation is constructed. Section 3 will show that the time fractional Schrödinger equation is equivalent to the usual Schrödinger equation but with a time dependent Hamiltonian. The solutions will not be invariant under time reversal nor will probability be conserved. In Sec. 4 solutions are obtained for a free particle. In Sec. 5 a solution for a potential well is found. In both cases it will be shown that probability increases over time to a limiting value that depends on the order of the time derivative. Section 6 presents some properties for the case of the order of the time derivative being between one and two. Concluding remarks are given in Sec. 7. There are also two appendices. The first is a primer on Caputo fractional calculus and a rationale of why it should be used to study physical systems instead of Riemann-Liouville fractional calculus. A second appendix presents some properties of the Mittag-Leffler function along with a new identity.

## II. TIME FRACTIONAL SCHRÖDINGER EQUATION

To fractionalize the time derivative of the Schrödinger equation care must be taken to preserve the units of the wave function. To begin with, note the definitions and the relationships of the Plank units given below (Plank length, time, mass, and energy)[12]:

$$L_p = \sqrt{\frac{Gh}{c^3}}, \quad T_p = \sqrt{\frac{Gh}{c^5}}, \quad M_p = \sqrt{\frac{hc}{G}}, \quad E_p = M_p c^2. \tag{5}$$

The standard form of the Schrödinger equation in one space and one time dimension is[13]

$$i\hbar \partial_t \psi = -\frac{\hbar^2}{2m} \partial_x^2 \psi + V\psi. \tag{6}$$

This can be expressed in Plank units as follows:

$$iT_p \partial_t \psi = -\frac{L_p^2 M_p}{2m} \partial_x^2 \psi + \frac{V}{E_p} \psi. \tag{7}$$

Define two dimensionless parameters; $N_m = m/M_p$ the number of Plank masses in $m$ and $N_V = V/E_p$ the number of Plank energies in $V$

$$iT_p \partial_t \psi = -\frac{L_p^2}{2N_m} \partial_x^2 \psi + N_V \psi. \tag{8}$$

When the time derivative is fractionalized there are two options,

$$(iT_p)^\nu D_t^\nu \psi = -\frac{L_p^2}{2N_m} \partial_x^2 \psi + N_V \psi, \tag{9}$$

$$i(T_p)^\nu D_t^\nu \psi = -\frac{L_p^2}{2N_m} \partial_x^2 \psi + N_V \psi. \tag{10}$$

In this paper, $D_t^\nu$ shall denote the Caputo fractional derivative of order $\nu$. $T_p$ must be raised to the same order as the fractional derivative to preserve the units of $\psi$.

The question of whether or not to raise *i* to the order of the time derivative needs more investigation. There are two reasons to choose Eq. (9) over Eq. (10), one superficial and one physical. When performing a Wick rotation the imaginary unit is raised to the same power as the time coordinate. The second reason involves the temporal behavior of the solution. When solving for the time component of Eq. (9) or Eq. (10) the Laplace transform is the preferred method. For Eq. (9), changing the order of the derivative moves the pole (from the inverse Laplace transform) up or down the negative imaginary axis. Hence, the temporal behavior of the solution will not change. For Eq. (10), changing the order of the derivative moves the pole to almost any desired location in the complex plane. Physically, this would mean that a small change in the order of the time derivative, in Eq. (10), could change the temporal behavior from sinusoidal to growth or to decay. Due to the simpler physical behavior of Eq. (9) and the role of '*i*' in a Wick rotation, Eq. (9) is the best candidate for a time fractional Schrödinger equation.

## III. TIME DEPENDENT HAMILTONIAN AND PROBABILITY CURRENT

The order of the time derivative on the left hand side Eq. (9) is not one, therefore, the operator on the right hand side is not a Hermitian Hamiltonian. Equation (5) can be rearranged to expose a time dependent Hamiltonian. To do so, the order of the time derivative on the left hand side of Eq. (9) must be raised to order one. First note an identity for Caputo derivatives for $0 < \nu < 1$,

$$D_t^{1-v} D_t^v y(t) = \frac{d}{dt} y(t) - \frac{\left[D_t^v y(t)\right]_{t=0}}{t^{1-v} \Gamma(v)}. \tag{11}$$

Now define two parameters:

$$\alpha = \frac{N_V}{T_p^v}, \tag{12}$$

$$\beta = \frac{(L_p)^2}{2 N_m (T_p)^v}, \tag{13}$$

then Eq. (9) can be written as

$$D_t^v \psi = -\frac{\beta}{i^v} \partial_x^2 \psi + \frac{\alpha}{i^v} \psi. \tag{14}$$

Using the identity for Caputo derivatives, Eq. (11), the time dependent Hamiltonian appears

$$\partial_t \psi = -\frac{\beta}{i^v} \partial_x^2 \left(D_t^{1-v} \psi\right) + \frac{\alpha}{i^v} \left(D_t^{1-v} \psi\right) + \frac{\left[D_t^v \psi(t)\right]_{t=0}}{t^{1-v} \Gamma(v)}. \tag{15}$$

Since the Hamiltonian is time dependent we should not expect probability to be conserved. Note also that the Hamiltonian is non-local in time, due to the integral in the formulation of the fractional derivative (see the appendix). This non-local character tells us that the solutions will not be invariant under time reversal. The third term in the Hamiltonian goes to zero as time goes to infinity (recall that the order of the original fractional time derivative is less than one).

Consider the non-local object in Eq. (15)

$$D_t^{1-v} \psi(t, x) = \frac{1}{\Gamma(1-v)} \int_0^t \left(\frac{d}{d\tau} \psi(\tau, x)\right) \frac{d\tau}{(t-\tau)^v}. \tag{16}$$

For a possible interpretation of this object recall the interpretation of the first order time derivative in standard quantum mechanics,[13)]

$$\frac{\partial}{\partial t} = \frac{E}{i\hbar}, \qquad (17)$$

where $E$ is viewed as the energy operator (Hamiltonian). Then the inner product (* denotes complex conjugation),

$$\int_{-\infty}^{\infty} \psi(t,x)^* D_t^{1-\nu} \psi(t,x)\, dx, \qquad (18)$$

can be interpreted as being proportional to the weighted time average of the energy of the wave function, the weighting factor being $(t-\tau)^{-\nu}$. For the remainder of the paper denote $\tilde{\psi} = D_t^{1-\nu}\psi$.

The probability current equation (for a free particle) can be constructed just as for the non-fractional Schrödinger equation.

$$probability\ density\ =\ P\ =\ \psi\psi^*, \qquad (19)$$

$$\partial_t P\ =\ \partial_t\psi\psi^*\ +\ \psi\partial_t\psi^*, \qquad (20)$$

$$\partial_t P\ =\ \left(-\frac{\beta}{i^\nu}\partial_x^2\tilde{\psi} + \frac{[D_t^\nu\psi(t,x)]_{t=0}}{t^{1-\nu}\Gamma(\nu)}\right)\psi^*\ +\ \psi\left(-\frac{\beta}{(-i)^\nu}\partial_x^2\tilde{\psi}^* + \frac{[D_t^\nu\psi(t,x)^*]_{t=0}}{t^{1-\nu}\Gamma(\nu)}\right). \qquad (21)$$

This can be rearranged

$$\partial_t P + \beta\partial_x\left(\frac{\partial_x\tilde{\psi}\psi^*}{i^\nu} + \frac{\partial_x\tilde{\psi}^*\psi}{(-i)^\nu}\right) = \beta\left(\frac{\partial_x\tilde{\psi}\partial_x\psi^*}{i^\nu} + \frac{\partial_x\tilde{\psi}^*\partial_x\psi}{(-i)^\nu}\right) + \frac{\psi^*[D_t^\nu\psi(t,x)]_{t=0} + \psi[D_t^\nu\psi(t,x)^*]_{t=0}}{t^{1-\nu}\Gamma(\nu)}. \qquad (22)$$

If the Hamiltonian were not time dependent (i.e. if $\nu \to 1$) then the right hand side of Eq. (22) would be zero. The probability current can be identified as

$$J = \frac{\beta}{i^\nu}(\partial_x\tilde{\psi})\psi^* + \frac{\beta}{(-i)^\nu}\psi(\partial_x\tilde{\psi}^*). \qquad (23)$$

The right hand side of Eq. (22) can be viewed as a source in the probability current equation. The non-zero source term in Eq. (22) confirms that probability will not be preserved for solutions of the time fractional Schrödinger equation.

$$S(x,t) = \frac{\beta}{i^\nu} \partial_x \tilde{\psi} \partial_x \psi^* + \frac{\beta}{(-i)^\nu} \partial_x \tilde{\psi}^* \partial_x \psi + \frac{\psi^*\left[D_t^\nu \psi(t,x)\right]_{t=0} + \psi\left[D^\nu \psi(t,x)^*\right]_{t=0}}{t^{1-\nu}\Gamma(\nu)}, \quad (24)$$

$$\partial_t P + \partial_x J = S. \quad (25)$$

Integrating Eq. (25) over all space and requiring the wave function and its first derivative go to zero at spatial infinity gives

$$\partial_t \int_{-\infty}^{\infty} P \, dx = \int_{-\infty}^{\infty} S \, dx. \quad (26)$$

## IV. FREE PARTICLE SOLUTION

The time fractional Schrödinger equation for a free particle is given by

$$(iT_p)^\nu D_t^\nu \psi = -\frac{L_p^2}{2N_m} \partial_x^2 \psi. \quad (27)$$

To solve this equation, apply a Fourier transform on the spatial coordinate, $F(\psi(x,t)) = \Psi(\lambda,t)$

$$(iT_p)^\nu D_t^\nu \Psi = \frac{(L_p \lambda)^2}{2N_m} \Psi. \quad (28)$$

The resulting equation can be rearranged and the results of the second appendix can be used. Namely, the identity for the Mittag-Leffler function with a complex argument

$$D_t^\nu \Psi = \frac{(L_p \lambda)^2}{2N_m T_p^\nu i^\nu} \Psi, \quad (29)$$

$$\Psi = \Psi_0 E_\nu\left(\omega(-it)^\nu\right), \tag{30}$$

or
$$\Psi = \frac{\Psi_0}{\nu}\left\{e^{-i\omega^{1/\nu} t} - \nu F_\nu\left(\omega(-i)^\nu, t\right)\right\}, \tag{31}$$

where $\omega = \frac{(L_p \lambda)^2}{2 N_m T_p^\nu}$. In Eq. (31) the first term is oscillatory, and the second is decay in the time variable. The function, $F_\nu$, is defined in the second appendix. Inverse Fourier transforming gives the final solution

$$\psi(x,t) = F^{-1}\Psi(\lambda,t) = \frac{1}{2\pi}\int_{-\infty}^{\infty} e^{i\lambda x} \frac{\Psi_0}{\nu}\left\{e^{-i\omega^{1/\nu} t} - \nu F_\nu\left(\omega(-i)^\nu, t\right)\right\} d\lambda. \tag{32}$$

This can be broken into two parts, a Schrödinger like piece divided by $\nu$, and a decay term that goes to zero as time goes to infinity

$$\psi_S(x,t) = \frac{1}{2\pi\nu}\int_{-\infty}^{\infty} e^{i\lambda x} \Psi_0 e^{-i\omega^{1/\nu} t} d\lambda, \tag{33}$$

$$\psi_D(x,t) = \frac{-1}{2\pi}\int_{-\infty}^{\infty} e^{i\lambda x} \Psi_0 F_\nu\left(\omega(-i)^\nu, t\right) d\lambda, \tag{34}$$

$$\psi(x,t) = \psi_S(x,t) + \psi_D(x,t). \tag{35}$$

Note that as $\nu$ goes to one the decay term goes to zero and the Schrödinger like term becomes the non-fractional Schrödinger term.

$\Psi_0$ may be chosen so that the initial probability is one

$$\int_{-\infty}^{\infty} \psi(x,0)\psi^*(x,0) dx = 1. \tag{36}$$

Due to the decay term in the solution one may ask what happens to the total probability as time goes to infinity. Consider the following limit

$$\lim_{t \to \infty} \int_{-\infty}^{\infty} \psi(x,t)\psi^*(x,t) dx \tag{37}$$

$$= \lim_{t \to \infty} \int_{-\infty}^{\infty} F^{-1}\left(\frac{\Psi_0}{\nu}\left\{e^{-i\omega^{1/\nu}t} - \nu F_\nu\left(\omega(-i)^\nu, t\right)\right\}\right) F^{-1}\left(\frac{\Psi_0}{\nu}\left\{e^{-i\omega^{1/\nu}t} - \nu F_\nu\left(\omega(-i)^\nu, t\right)\right\}\right)^* dx. \quad (38)$$

Now use Parseval's identity

$$= \frac{2\pi}{\nu^2} \lim_{t \to \infty} \int_{-\infty}^{\infty} \Psi_0\left\{e^{-i\omega^{1/\nu}t} - \nu F_\nu\left(\omega(-i)^\nu, t\right)\right\}\left(\Psi_0\left\{e^{-i\omega^{1/\nu}t} - \nu F_\nu\left(\omega(-i)^\nu, t\right)\right\}\right)^* d\lambda. \quad (39)$$

In the limit that time goes to infinity $F_\nu$ goes to zero, this leaves

$$= \frac{2\pi}{\nu^2} \lim_{t \to \infty} \int_{-\infty}^{\infty} \Psi_0 e^{-i\omega^{1/\nu}t} \Psi_0^* e^{i\omega^{1/\nu}t} d\lambda, \quad (40)$$

$$= \frac{2\pi}{\nu^2} \lim_{t \to \infty} \int_{-\infty}^{\infty} \Psi_0 \Psi_0^* d\lambda. \quad (41)$$

Now use Parseval's identity again

$$= \frac{1}{\nu^2} \lim_{t \to \infty} \int_{-\infty}^{\infty} \psi_0 \psi_0^* dx. \quad (42)$$

The remaining integral is unity. Hence,

$$\lim_{t \to \infty} \int_{-\infty}^{\infty} \psi(x,t) \overline{\psi}(x,t) dx = \frac{1}{\nu^2}. \quad (43)$$

Since $\nu$ is less than one, the total probability increases over time to the limiting value of Eq. (43).

## V. POTENTIAL WELL SOLUTION

Now consider a particle in a potential well

$$V(x) = \begin{cases} 0 & 0 < x < a \\ \infty & \text{elsewhere} \end{cases}, \quad (44)$$

$$\left(iT_p\right)^\nu D_t^\nu \psi = -\frac{L_p^2}{2N_m} \partial_x^2 \psi, \quad (45)$$

$$\psi(0,t) = 0,$$
$$\psi(a,t) = 0. \tag{46}$$

This can be solved by separation of variables, $\psi = A(t)B(x)$

$$(iT_p)^\nu \frac{D_t^\nu A}{A} = -\frac{L_p^2}{2N_m} \frac{\partial_x^2 B}{B} = \lambda. \tag{47}$$

Solve the spatial component first

$$B'' + \lambda \frac{2N_m}{L_p^2} B = 0. \tag{48}$$

The boundary conditions give

$$B_n = c_n \sin\left(\frac{n\pi}{a} x\right), \tag{49}$$

$$\lambda_n = \left(\frac{n\pi L_p}{a}\right)^2 \frac{1}{2N_m}. \tag{50}$$

The time equation is

$$D_t^\nu A = \frac{\lambda_n}{(iT_p)^\nu} A. \tag{51}$$

This can be solved in terms of the Mittag-Leffler function (see Appendix B). Let $A(t = 0) = 1$ so that we can specify an initial wave function.

$$A = E_\nu\left(\omega_n(-it)^\nu\right), \tag{52}$$

$$A = \frac{1}{\nu}\left\{e^{-i\omega^{1/\nu} t} - \nu F_\nu\left((-i\omega)^\nu, t\right)\right\}, \tag{53}$$

where, $\omega_n = \dfrac{\lambda_n}{T_p^\nu}$.

Note also that

$$\lim_{t \to \infty} |A(t)| = \frac{1}{\nu}. \tag{54}$$

The normalized spatial eigenfunctions are

$$\psi_n(x,0) = \sqrt{\frac{2}{a}} \sin(n\pi x/a), \tag{55}$$

$$\int_0^a \psi_n(x,0) \cdot \overline{\psi}_n(x,0) dx = 1. \tag{56}$$

The solutions for all times can be written as

$$\psi_n(x,t) = \sqrt{\frac{2}{a}} \sin(n\pi x/a) \frac{1}{\nu} \left\{ e^{-i\omega^{1/\nu} t} - \nu F_\nu \left( (-i\omega)^\nu, t \right) \right\}. \tag{57}$$

As in the free particle case it is interesting to compute the limit of the total probability as time goes to infinity

$$\lim_{t \to \infty} \int_0^a \psi_n(x,t) \cdot \overline{\psi}_n(x,t) dx = \frac{1}{\nu^2}. \tag{58}$$

This is the same result as for the free particle. As $\nu$ is less than one, the total probability is greater than one as time goes to infinity. In fact as long as $t_1 < t_2$ we have

$$\int_0^a \psi(x,t_1) \cdot \overline{\psi}(x,t_1) dx < \int_0^a \psi(x,t_2) \cdot \overline{\psi}(x,t_2) dx. \tag{59}$$

Probability is created as time progresses. This can also be viewed as particles are created (extracted from the potential) as time progresses.

The energy levels for the potential well can be computed. Since the Hamiltonian is time dependent the energy levels will also be time dependent.

$$E_n(t) = \int_0^a \psi^* i\hbar \partial_t \psi \, dx, \tag{60}$$

$$E_n(t) = \frac{2i\hbar}{a\nu^2} \left\{ e^{i\omega^{1/\nu} t} - \nu F_\nu \left( (i\omega)^\nu, t \right) \right\} \partial_t \left\{ e^{-i\omega^{1/\nu} t} - \nu F_\nu \left( (-i\omega)^\nu, t \right) \right\} \int_0^a \sin^2\left( \frac{n\pi x}{a} \right) dx, \tag{61}$$

$$E_n(t) = \frac{i\hbar}{v^2}\left\{e^{i\omega^{1/v}t} - v F_v\left((i\omega)^v,t\right)\right\}\partial_t\left\{e^{-i\omega^{1/v}t} - v F_v\left((-i\omega)^v,t\right)\right\}. \tag{62}$$

The interesting result is when time goes to infinity

$$E_n(\infty) = \frac{\hbar(\lambda_n)^{1/v}}{v^2 T_p} = \frac{\hbar(n\pi L_p)^{2/v}}{v^2 T_p (2a^2 N_m)^{1/v}}. \tag{63}$$

This is the same energy spectrum that is obtained for the non-fractional Schrödinger equation except for the factor of $1/v^2$ and the exponent of $1/v$. The difference between two energy levels, say $m$ and $n$, is given by

$$E_m(\infty) - E_n(\infty) = \frac{\hbar(\pi L_p)^{2/v}}{v^2 T_p (2a^2 N_m)^{1/v}}\left(m^{2/v} - n^{2/v}\right). \tag{64}$$

Since $v$ is less than one the spacing between energy levels is greater than that given by the non-fractional Schrödinger equation. In fact the smaller the value of $v$ the greater the difference between energy levels. Note also that at $t = 0$ the spacing between the energy levels is that same as that of the non-fractional Schrödinger equation. As time progresses the spacing between the energy levels increases to the limiting value given by Eq. (64). Hence, radiation that is emitted, say from state $n$ to state $m$, at an early time will have a longer wavelength than radiation emitted at a later time.

## VI. SOME PROPERTIES FOR $1 < v \leq 2$

In this section the case of $1 < v \leq 2$ is briefly considered. Notice that at the upper limit we have a special case of the Klein-Gordon equation. Like the Klein-Gordon equation, the initial value of the first derivative must also be specified to obtain a solution. Just as in the previous case, Eq. (14) can be recast to expose a time dependent

Hamiltonian. First note an identity for Caputo fractional derivatives and integrals for $1 < \nu \le 2$

$$I_t^{\nu-1} D_t^\nu f(t) = D_t^1 f(t) - D_t^1 f(t)\big|_{t=0}. \tag{65}$$

Applying $I_t^{\nu-1}$ to Eq. (14) yields

$$\partial_t \psi = -\frac{\beta}{i^\nu} \partial_x^2 \left( I_t^{1-\nu} \psi \right) + \frac{\alpha}{i^\nu} \left( I_t^{1-\nu} \psi \right) + \partial_t \psi(t)\big|_{t=0}. \tag{66}$$

This is very similar to Eq. (15) except that the fractional derivatives are replaced with fractional integrals and the initial value term is for the first derivative.

The solution can also be written down for a free particle just like the previous case

$$(iT_p)^\nu D_t^\nu \psi = -\frac{L_p^2}{2N_m} \partial_x^2 \psi, \tag{67}$$

$$\psi(x,0) = \psi_0,$$

$$\frac{d}{dt} \psi(x,t)\bigg|_{t=0} = \psi_1.$$

Denote the Fourier transform (on the spatial coordinate) of the wave function and the initial conditions as

$$\begin{aligned} \Psi(\lambda,t) &= F(\psi(x,t)), \\ \Psi_0 &= F(\psi_0), \\ \Psi_1 &= F(\psi_1). \end{aligned} \tag{68}$$

The solution, in Fourier space, can then be expressed as

$$\Psi = \Psi_0 \left( \frac{e^{i\omega^{1/\nu} t}}{\nu} + F_\nu \left( \omega(-i)^\nu, t \right) \right) + \Psi_1 \left( \frac{e^{i\omega^{1/\nu} t}}{\omega^{1/\nu} \nu} - F_{\nu-1} \left( \omega(-i)^\nu, t \right) \right), \tag{69}$$

where, $\omega = \frac{(L_p \lambda)^2}{2 N_m T_p^\nu}$. This solution behaves just like the case of $0 < \nu \leq 1$ in that there is an oscillatory term and a decay term. The presence of a decay term will again cause probability to increase over time.

## VII. CONCLUSION

In this paper the time fractional Schrödinger equation was constructed. It was found to be equivalent to the usual Schrödinger equation but with a Hamiltonian that was time dependent and non-local in time. In contrast to the space fractional Schrödinger equation, probability was not conserved but found to increase over time to a limiting value depending on the order of the time derivative. Consequently, the energy of the eigenstates for the potential well were also found to increase over time to a limiting value (for $0 < \nu < 1$). In the limit that $t \to \infty$ the energy of the eigenstates was found to be similar to that given by the usual Schrödinger equation but divided by the order of the time derivative squared and having an exponent depending on the order of the time derivative. Initially the energy of the eigenstates is the same as for the non-fractional Schrödinger equation. As time progresses the energy of each level increases as does the spacing between the energy levels. Hence, if the spectrum of emitted radiation was monitored over time, for such a fractional well, it would be seen to blue shift and, the space between the spectral lines would also increase. In finding a solution to the time fractional Schrödinger equation a new identity for the Mittag-Leffler function was found. This identity generalizes the Euler identity for the exponential function with complex arguments.

**ACKNOWLEDGEMENT**

The author would like to thank Jean Krisch, Ed Glass, and David Garfinkle for many useful and stimulating discussions. The author would also like to thank P. Dorcey for critically reading an early version of the paper. The referee is also thanked for several useful suggestions.

**APPENDIX A: CAPUTO FRACTIONAL CALCULUS**

The bulk of this appendix is taken from Ref. 14. The two most commonly used definitions of fractional derivatives are the Riemann-Liouville and Caputo (there are many definitions, the reader is encouraged to consult Ref. 15 for a more complete discussion). Each definition uses Riemann-Liouville fractional integration and derivatives of whole order. The difference between the two definitions is in the order of evaluation. Riemann-Liouville fractional integration of order $\mu$ is defined as

$$\mathrm{I}^{\mu}(f(t)) = \frac{1}{\Gamma(\mu)} \int_0^t \frac{f(\tau)\mathrm{d}\tau}{(t-\tau)^{1-\mu}}. \tag{A1}$$

The next two equations define Riemann-Liouville and Caputo fractional derivatives of order $\nu$, respectively,

$$^{RL}\mathrm{D}_t^{\nu} f(t) = \frac{\mathrm{d}^k}{\mathrm{d}t^k}\left(\mathrm{I}^{k-\nu} f(t)\right), \tag{A2}$$

$$^{C}\mathrm{D}_t^{\nu} f(t) = \mathrm{I}^{k-\nu}\left(\frac{\mathrm{d}^k}{\mathrm{d}t^k} f(t)\right), \tag{A3}$$

where $k-1 \leq \nu < k$. For now, the Caputo fractional derivative will be denoted by $^{C}\mathrm{D}_t^{\nu}$ to maintain a clear distinction with the Riemann-Liouville fractional derivative. The Caputo fractional derivative first computes an ordinary derivative followed by a fractional integral to achieve the desired order of fractional derivative. The Riemann-Liouville fractional derivative is computed in the reverse order.

The desire to formulate initial value problems for physical systems leads to the use of Caputo fractional derivatives rather than Riemann-Liouville fractional derivatives. Consider the Laplace transform of the Riemann-Liouville fractional derivative,

$$L_t\{^{RL}D_t^\nu f(t)\} = s^\nu F(s) - \sum_{k=0}^{n-1} s^k \left(^{RL}D_t^{\nu-k-1} f(t)\right)\bigg|_{t=0}. \tag{A4}$$

The initial conditions, $\left(^{RL}D_t^{\nu-k-1} f(t)\right)\big|_{t=0}$ for $k = 0, \mathrm{K}, n-1$, are fractional order derivatives (see Ref. 15 for a detailed discussion of these objects). When studying a physical system, initial conditions are typically conditions that can be measured or imposed on the system. As yet there is no physical interpretation for $\left(^{RL}D_t^{\nu-k-1} f(t)\right)\big|_{t=0}$ (see Refs. 16 and 17 for a fractal interpretation of the fractional order integral). Some authors using equations with Riemann-Liouville fractional derivatives to model physical systems have added terms to the diffusion equation to eliminate these unphysical terms (see, e.g., Ref. 18).

The Laplace transform of the Caputo fractional derivative is

$$L_t\{^{C}D_t^\nu f(t)\} = s^\nu F(s) - \sum_{k=0}^{n-1} s^{\nu-k-1} \left(D_t^k f(t)\right)\bigg|_{t=0}. \tag{A5}$$

In this case the initial conditions are well understood from a physical point of view. For example if $f(t)$ represents position then $\left(D_t^0 f(t)\right)\big|_{t=0}$ is the initial position, $\left(D_t^1 f(t)\right)\big|_{t=0}$ is the initial velocity, etc. (see Podlubny[15] section 2.4 for a more detailed discussion of this point).

## APPENDIX B: THE MITTAG-LEFFLER FUNCTION

The Mittag-Leffler function, $E_v(t)$, is a generalization of the exponential function (see Refs. 15 and 19 for additional properties and a history)

$$e^t = \sum_{n=0}^{\infty} \frac{t^n}{n!} = \sum_{n=0}^{\infty} \frac{t^n}{\Gamma(n+1)}, \tag{B1}$$

$$E_v(t) = \sum_{n=0}^{\infty} \frac{t^n}{\Gamma(vn+1)}. \tag{B2}$$

One very useful formula concerning the exponential function is with a complex argument, the Euler identity

$$e^{it} = \cos(t) + i\sin(t). \tag{B3}$$

A similar identity occurs for the Mittag-Leffler function. The derivation is given below. Consider the initial value fractional differential equation

$$\begin{aligned} {_0D_t^v} A &= \sigma A, \\ A(t=0) &= A_0. \end{aligned} \tag{B4}$$

The solution of this can be found by Laplace transform in two different ways

$$s^v \tilde{A} - s^{v-1} A_0 = \sigma \tilde{A}, \tag{B5}$$

$$\tilde{A} = \frac{s^{v-1} A_0}{s^v - \sigma}. \tag{B6}$$

Equation (B6) can be expressed as a power series

$$\tilde{A} = A_0 \sum_{n=0}^{\infty} \frac{\sigma^n}{s^{1+vn}}. \tag{B7}$$

Inverting this series yields a series representation for the Mittag-Leffler function

$$A = A_0 \sum_{n=0}^{\infty} \frac{\sigma^n t^{vn}}{\Gamma(vn+1)} = A_0 E_v(\sigma t^v). \tag{B8}$$

The solution to

$$_0D_t^\gamma A = \sigma i^v A,$$
$$A(t = 0) = A_0, \tag{B9}$$

is then given by

$$A = A_0 E_v\left(\sigma i^v t^v\right). \tag{B10}$$

To see how this breaks up into an oscillatory piece and a decay piece consider the Laplace transform of Eq. (B9)

$$\tilde{A} = \frac{s^{v-1} A_0}{s^v - \sigma i^v}. \tag{B11}$$

The inverse Laplace transform is given by

$$A(t) = \frac{1}{2\pi i} \int_{\gamma - i\infty}^{\gamma + i\infty} \frac{e^{st} s^{v-1} A_0}{s^v - \sigma i^v} ds. \tag{B12}$$

The inverse Laplace transform can be evaluated using residues. This integral has a branch point at $s = 0$ due to $s^{v-1}$ in the numerator and $s^\mu$ in the denominator, and a pole at $s_0 = \sigma^{1/v} i$. Due to the branch point at the origin the usual Bromwich contour cannot be used. A branch cut along the negative Real($s$) axis must be made. That is, a cut from $-\infty$ into and then around the origin in a clockwise sense and then back out to $-\infty$. The usual Bromwich contour is continued after the cut. This is referred to as a Hankel contour. The solution will then be given by the residue minus the contribution along the partial path

$$A(t) = \text{Residue} - \frac{A_0}{2\pi i} \int_{Hankel}. \tag{B14}$$

The residue is given by

$$\frac{\text{Residue}}{A_0} = \lim_{s \to s_0} \frac{(s-s_0)e^{st}s^{\nu-1}}{s^\nu - s_0^\nu} = \frac{e^{i\sigma^{1/\nu}t}}{\nu}. \tag{B15}$$

The integral along the Hankel contour only makes contributions along the branch cut. This contribution is given by

$$-\frac{A_0 \sigma i^\nu}{\pi} \int_0^\infty \frac{\sin(\nu\pi)e^{-rt}r^{\nu-1}dr}{r^{2\nu} - 2\sigma i^\nu \cos(\nu\pi)r^\nu + (\sigma i^\nu)^2}. \tag{B16}$$

Combining the path and residue contributions, and doing some algebra gives

$$\frac{A}{A_0} = \frac{e^{i\sigma^{1/\nu}t}}{\nu} - \frac{\sigma i^\nu \sin(\nu\pi)}{\pi} \int_0^\infty \frac{e^{-rt}r^{\nu-1}dr}{r^{2\nu} - 2\sigma i^\nu \cos(\nu\pi)r^\nu + (\sigma i^\nu)^2}. \tag{B17}$$

The first term is oscillatory, and the second decays monotonically in the time variable. Hence, the Mittage-Leffler function, with a complex argument, can be expressed as

$$E_\nu(\sigma i^\nu t^\nu) = \frac{e^{i\sigma^{1/\nu}t}}{\nu} - \frac{\sigma i^\nu \sin(\nu\pi)}{\pi} \int_0^\infty \frac{e^{-rt}r^{\nu-1}dr}{r^{2\nu} - 2\sigma i^\nu \cos(\nu\pi)r^\nu + (\sigma i^\nu)^2}. \tag{B18}$$

Note that as $\nu \to 1$ the above equation becomes the Euler identity.

To make this result appear more compact define a new function

$$F_\nu(\rho,t) = \frac{\rho \sin(\nu\pi)}{\pi} \int_0^\infty \frac{e^{-rt}r^{\nu-1}dr}{r^{2\nu} - 2\rho\cos(\nu\pi)r^\nu + \rho^2}. \tag{B19}$$

This function decays monotonically in time. The following results are special cases:

$$F_\nu(0,t) = 0, \tag{B20}$$

$$F_1(\rho,t) = 0, \tag{B21}$$

$$F_\nu(\rho,0) = \frac{1-\nu}{\nu}, \tag{B22}$$

$$0 \leq F_\nu(\rho,t) \leq \frac{1-\nu}{\nu}. \tag{B23}$$

Hence, a solution to $_0D_t^\nu A = \sigma i^\upsilon A$, $A(t=0) = A_0$ can be written as

$$\frac{A}{A_0} = \frac{e^{i\sigma^{1/\nu} t}}{\nu} - F_\nu\left(\sigma i^\nu, t\right).$$ (B24)

Additionally, the solution to $_0D_t^\nu A = \sigma(-i)^\upsilon A$, $A(t=0) = A_0$ is given by

$$\frac{A}{A_0} = \frac{e^{-i\sigma^{1/\nu} t}}{\nu} - F_\nu\left(\frac{\sigma}{i^\nu}, t\right).$$ (B25)

To recap, the "Euler identity" for the Mittag-Leffler function is

$$E_\nu\left(\sigma i^\nu t^\nu\right) = \frac{e^{i\sigma^{1/\nu} t}}{\nu} - F_\nu\left(\sigma i^\nu, t\right).$$ (B26)

In a similar fashion, a solution can be worked out for $1 < \nu < 2$

$$\begin{aligned} _0D_t^\nu A &= \sigma i^\upsilon A, \\ A(t=0) &= A_0, \\ _0D_t^1 A(t=0) &= A_1, \end{aligned}$$ (B27)

$$A = A_0\left(\frac{e^{i\sigma^{1/\nu} t}}{\nu} + F_\nu\left(\sigma i^\nu, t\right)\right) + A_1\left(\frac{e^{i\sigma^{1/\nu} t}}{\sigma^{1/\nu}\nu} - F_{\nu-1}\left(\sigma i^\nu, t\right)\right).$$ (B28)